\documentclass[referee]{raa}    

\usepackage{graphicx,times}
\usepackage{natbib}
\usepackage{amssymb,amsmath}
\bibpunct{(}{)}{;}{a}{}{,}

\begin{document}

   \title{Probing the wind and evolutionary state of the quiescent luminous blue variable Var\,2 in the Galaxy M33: constraints from self-consistent non-LTE modeling}

 \volnopage{ {\bf 20XX} Vol.\ {\bf X} No. {\bf XX}, 000--000}
   \setcounter{page}{1}

   \author{A.~Kostenkov
      \inst{1}
   \and Y.~Solovyeva
      \inst{1}
   \and E.~Dedov
      \inst{1}
   \and A.~Vinokurov
      \inst{1}
   \and A.~Sarkisyan
      \inst{1}
   }

   \institute{$^{1}$ Special Astrophysical Observatory, Nizhnij Arkhyz 369167, Russia; {\it kostenkov@sao.ru}\\
\vs \no
   {\small Received 20XX Month Day; accepted 20XX Month Day}
}

\abstract{ Luminous Blue Variables (LBVs) are massive stars with extreme luminosities, exhibiting significant irregular photometric and spectroscopic variability. Key open questions regarding their defining characteristics and evolutionary role are complicated by the existence of quiescent LBVs that enter prolonged periods of stability. This paper is dedicated to the study of such dormant LBV Var\,2 in the galaxy M33, which has not shown significant brightness variability for almost century and currently observed as Ofpe\slash WN9 star. Our study aims to derive fundamental stellar and wind parameters of Var\,2, thereby constraining its current evolutionary state, initial mass and origin. In order to obtain reliable estimates of the fundamental parameters of Var\,2, we had calculated self-consistent hydrodynamic models of the extended atmosphere, assuming non-local thermodynamic equilibrium (non-LTE) and taking into account the balance of radial forces in the wind of the star. Spectral modeling yields luminosity $L=6.5\times10^{5}\,L_\odot$, current mass $M_* \approx23\,M_\odot$ and hydrogen abundance on the surface ${\rm X}_{\rm H}\approx43\%$. Comparison with evolutionary tracks indicates that these parameters correspond to a star with the initial mass $M_{\rm init}\approx50\,M_\odot$ and age $t_{\rm age}\approx4.4\,$Myr that is evolving from lower temperatures towards the Wolf-Rayet stars. A study of dynamic properties of the wind showed that the shape of the wind velocity profile of Var\,2 is close to the one of O- and early B-supergiants rather than Wolf-Rayet stars. In contrast, for the obtained mass-loss rate $\dot{M}=2.1\times10^{-5}\,M_{\odot}\,\text{yr}^{-1}$, the ratio of the wind momentum to the luminosity of Var\,2 is in good agreement with the values found for WNL stars. Given the obtained age estimates, Var\,2 could potentially have been ejected from the cluster associated with the nearest large star-forming region located at a distance of $\sim100\,$pc. However, statistical analysis of the projected distribution of stars in the vicinity of Var\,2 suggests that this LBV could have formed in a local low-populated group.
}

   \authorrunning{A.~Kostenkov et al.}            %author_head in even pages
   \titlerunning{The quiescent luminous blue variable Var\,2 in the Galaxy M33}  % title_head in odd pages
   \maketitle

%________________________________________________ sections below
% 
\section{Introduction}           %% first-level sections will be auto-capitalized

Research on massive stars plays a crucial role in understanding the evolution of the interstellar medium and galaxies. These objects are the main source of both ionizing photons and heavy elements ejected into the surrounding environment with a number of physical phenomena (e.g. stellar wind or supernova explosions). The combination of the complexity of physical processes that determine the observed properties of massive stars with the observed rarity of such objects leads to several uncertainties regarding the evolutionary status as well as stellar parameters determination problems.

The short-term ($\sim10^3-10^5$ years) post-Main Sequence stages of the evolution of the most massive stars ($M_{\rm init}>20\,M_\odot$) are of particlar interest. These stages are known to be accompanied by the substantial mass-loss, which determines their further evolution \citep{Humphreys1994, deJager1998, Smith2014}. Luminous blue variables (LBVs) serve as one example of the objects in a such phase of the evolutionary path. Observational properties of LBVs are characterized by simultaneous irregular photometric and spectral variability, and in some cases spontaneous outbursts \citep{vanGenderen2001, Weis2020}. In contrast, some LBVs have remained stable for decades following an active phase. A prime example of such a "dormant" LBV is the Galactic LBV P\,Cygni, which erupted in the 17th century and has remained relatively stable for centuries with a moderate brightness increase of 0.15\,mag per 100 years \citep{Lamers1992, Elliott2022}. This prolonged quiescence sometimes precedes a renewed activity \citep{Polcaro2016, Smith2020}, but the physical link between the two states remains unknown \citep{Maryeva2022}.

In the classical theory of massive star evolution, LBVs corresponds to the transition from hydrogen burning to helium burning in the core, accompanied by intense mass outflow and episodic massive ejections of matter that lead to the loss of the outer hydrogen layers (the Conti scenario, \cite{Conti1975, Conti1984, Humphreys1994}). Thus, LBVs are presumably intermediate in evolution between massive O-type stars ($M_{\rm init}\gtrsim$ 20-25 M$_\odot$) and Wolf-Rayet (WR) stars \citep{Meynet2011, Weis2020, Maryeva2024}. At the same time, some data indicate that the LBV phase corresponds to the final stage of the evolution of massive stars, ultimately preceding type IIn supernovae \citep{GalYam2009, Smith2010, Ustamujic2021}.

An alternative approach to the study of LBVs suggests that these objects are the product of the evolution of close binary systems with mass exchange \citep{KenyonGallagher1985, Justham2014, Smith2017}. One of the arguments in favor of this evolutionary scenario is the significant isolation of LBVs with respect to O-type stars in comparison with other massive stars \citep{Smith2015}. The proposed mechanisms for this isolation include mass transfer extending the lifetime of star to allow for migration, or binary disruption causing a high-velocity ejection \citep{Aghakhanloo2017}.

A critical step in testing different evolutionary scenarios is the accurate determination of fundamental stellar parameters, especially stellar mass. This task is particularly challenging for evolved massive stars with dense winds \citep{Sabhahit2025}. Such winds can completely obscure the inner photospheric layers, meaning the absorption spectral lines used for mass analysis are either absent or severely contaminated. A further complication arises from the poorly constrained dynamic properties of the winds, which determine the velocity profile and thus the structure of the pseudo-photosphere formed by the dense outflow.

Partially, these problems can be addressed by using self-consistent hydrodynamic wind models, assuming non-local thermodynamic equilibrium (non-LTE). Unlike standard non-LTE models that rely on an analytical wind velocity profile, which varies significantly in different studies (e.g. \citealt{Najarro1997, Groh2009, Maryeva2018}), these models solve for the balance of radial forces in the wind. However, substantial uncertainty persists in determining stellar parameters due to the inhomogeneous (clumpy) structure of stellar winds, which affects both their density and velocity profiles. This degeneracy can be resolved either through direct mass determination in binary systems \citep{Sabhahit2025} or by investigating clumps stratification via analysis of quasi-simultaneous observations across a wide spectral range \citep{Puls2006, Najarro2009, Najarro2011}. Currently, both approaches are feasible for only a handful of stars.

Thus far, hydrodynamic non-LTE models have been constructed primarily for blue super- and hypergiants \citep{Sander2017, GormazMatamala2021, BerniniPeron2025} or WR-stars \citep{GrafenerHamann2005, Grafener2008, Sabhahit2025}, while proper self-consistent non-LTE modeling for LBVs is still absent. A comparison of the derived model parameters with evolutionary tracks can improve understanding of the LBV phenomenon and refine current evolutionary models of the most massive stars. Furthermore, the obtained age estimate also can be an essential constraint for interpreting the local stellar environment.

This paper is devoted to the study of Var\,2 (V* Y Tri, J013418.35+303836.8), a confirmed LBV in the galaxy M33 (\cite{Richardson2018} and references therein). The variability of the object was discovered about 100 years ago \citep{Duncan1922, Wolf1922}. Var\,2 reached its maximum brightness of $m_{\rm pg}\approx16.5\,$mag in 1925, then over several decades a decrease in brightness down to $m_{\rm pg}\approx17.5\,$mag was observed. The star have appeared to be in a "dormant"{} state since 1940s, showing insignificant photometric variability \citep{HubbleSandage1953, Humphreys1975, Humphreys1978, Kurtev1999, Massey2007, Massey2016, Martin2017}. Var\,2 is currently observed as an Ofpe\slash WN9 star \citep{Neugent2011, Massey2016} with a low terminal wind velocity $v_\infty\approx230\,$km/s \citep{Humphreys2014}, typical of LBVs \citep{Groh2009, Najarro1997}. Notably, although the apparent magnitude $V=18.17$ mag and color indices $(U-B)=-1.0$, $(B-V)=-0.17$ given for Var\,2 by \cite{Humphreys1978} are close to modern ones, \cite{Humphreys1975, Humphreys1978} report that Var\,2 shows an A-F-type spectrum. This classification radically contradicts both the WNL-type spectrum currently observed for this object and its historical color measurements, which are typical for the hottest blue stars. Var\,2 is located approximately 35\arcsec{}{} from the nearest large group of bright blue stars \citep{Kostenkov2025, Burggraf2005}, however, it is unknown whether there is a link between them.

This work involves modeling of the optical spectrum and spectral energy distribution of Var\,2. We also perform a detailed investigation of its wind using self-consistent non-LTE calculations and compare the derived stellar parameters with evolutionary tracks. The paper is organized as follows: Section 2 presents the observational data used in the work and describes their processing. Section 3 is devoted to the determination of the fundamental stellar parameters of Var\,2 using modern non-LTE models of extended atmospheres. Section 4 presents a study of the Var\,2 wind dynamics with self-consistent radiation-hydrodynamic models of extended atmospheres. Section 5 discusses the obtained results in the context of the possible evolutionary status of Var\,2, considers the differences between the Var\,2 wind and the expanding envelopes of other massive stars and analyzes the nearest stellar environment of Var\,2.

\section{Observational data}
\label{obs_section}

\subsection{Spectral data}

\begin{figure*}
\centering
\includegraphics[scale=0.5, angle=0]{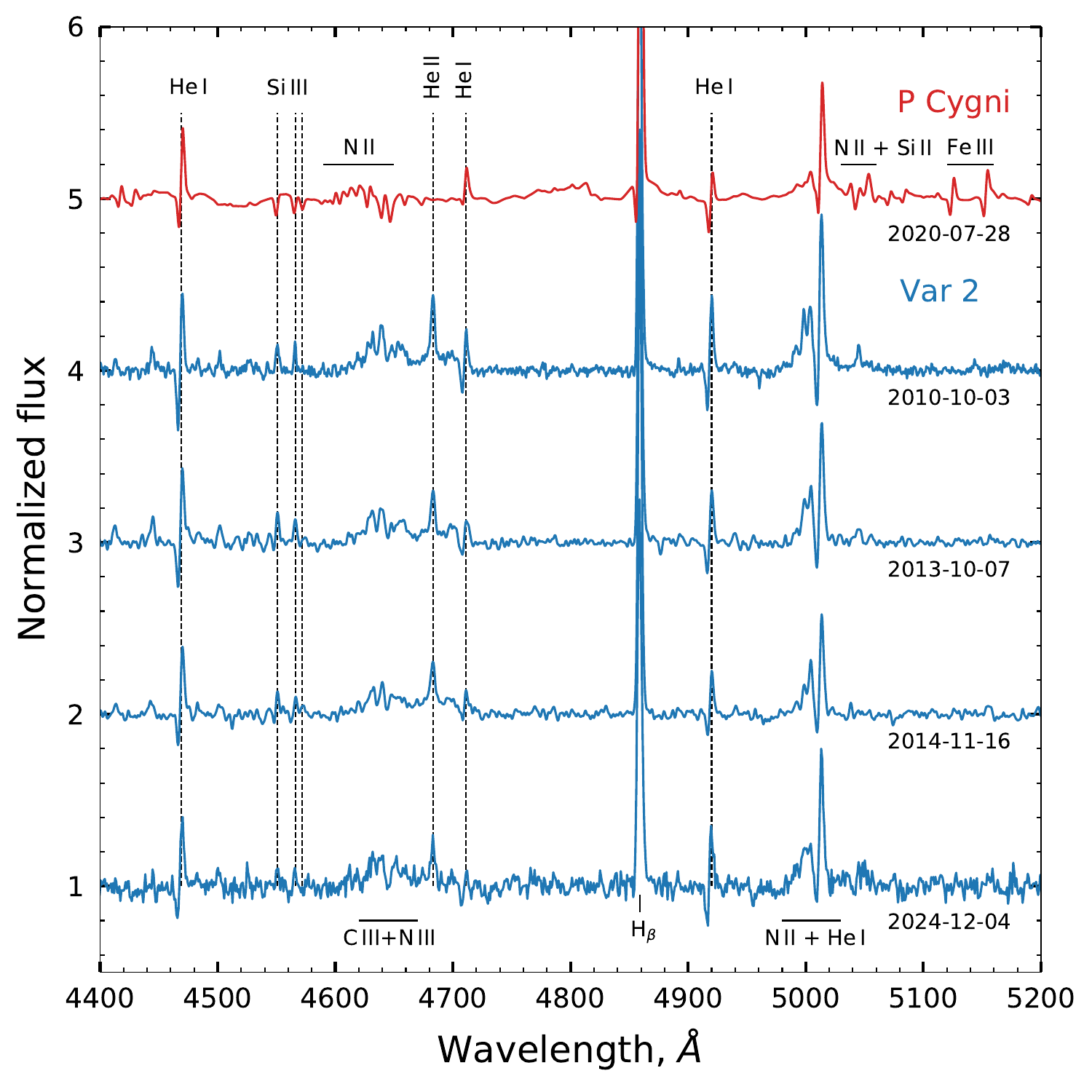}
\caption{The normalized optical spectra of Var\,2 obtained in different epochs (blue lines), and for comparison spectrum of P\,Cygni smoothed to 3\AA{} resolution (red line). The spectra are offset vertically for clarity. The normalized optical spectra of Var\,2 and P\,Cygni.}
\label{var2_pcygni}
\end{figure*}

The spectral data used for modeling were obtained on December 4, 2024 with the SCORPIO-1 focal reducer \citep{Afanasiev2005} mounted on the 6-meter Russian telescope. Observations were carried out using the VPHG1200B and VPHG1200R grisms, covering wavelength ranges of 3600-5400\,\AA{} and 5700-7500\,\AA, respectively. A slit width of 0.5\arcsec{}{} provided a spectral resolution of about 3\AA{} in both ranges, while the seeing remained stable at about 1.8\arcsec{}{} during spectroscopy. Data reduction was performed in the \textsc{midas} environment following standard long-slit spectral processing. One-dimensional spectra were extracted using the \textsc{spextra} software, specifically designed for crowded stellar fields \citep{Sarkisyan2017}. 

A comparison of the obtained spectrum with those presented in \cite{Humphreys2014, Humphreys2017}\footnote{Var\,2 spectra were taken from web archive of project "Eta Carinae and Giant Eruptions"{} (https://etacar.umn.edu/)} reveals no notable variability in the optical spectrum of Var\,2 over the past decade. Figure~\ref{var2_pcygni} presents the blue part of the optical spectra, which contains the highest concentration of diagnostic lines. Only minor variations in the equivalent widths of the He\,II, He\,I, and hydrogen lines are present.

It is likely that the last notable changes in the state of Var\,2 occurred in the late 20th century. In the 1970s, the star exhibited an anomalously cool spectrum despite showing extremely blue colors: $(U-B)=-1.0$ and $(B-V)=-0.17$ \citep{Humphreys1975, Humphreys1978}. Unfortunately, spectra of Var\,2 were not obtained from the late 1970s until the early 2010s, making it impossible to determine the exact timing of the transition from A-F to WNL-type spectrum. To address the unusual colors reported in earlier studies of Var\,2, one can assume that an extended atmosphere could significantly distort the observed colors in the blue wavelength range due to an inverse Balmer jump. Indeed, stars with dense outflows, that retain neutral hydrogen in their winds (including some LBV/cLBV, B[e]-supergiants, yellow hypergiants), show the $(U-B)$ color index typical of much hotter stars \citep{Massey2007}. However, the radical change in the spectral type of Var\,2, accompanied by a major shift in wind properties, would necessarily have been reflected in the photometry, as observed for LBVs \citep{Lamers1998, vanGenderen2001, Groh2009, Burggraf2015}. This is not the case for Var\,2, which had apparent magnitude and blue color indices during its potential cool phase that are close to those observed in the last decade. Therefore, this anomaly cannot be explained as an observational property of dense outflows, nor is it consistent with the photometric variability of other LBVs.

We compared the spectrum of Var\,2 with that of P\,Cygni (spectral type B1-2 Ia-0ep; \citealt{Skiff2014}), which is the prototype of "dormant"{} LBVs. For this purpose, we used an archival spectrum from the SOPHIE database\footnote{http://atlas.obs-hp.fr/sophie/}  \citep{Moultaka2004} obtained on July 28, 2020. As seen in Figure~\ref{var2_pcygni}, the spectrum of P\,Cygni is significantly cooler than those of Var\,2: it lacks the He\,II, C\,III, N\,III lines and shows a notable absorption component in the profile of the H$_\beta$ line.

\subsection{Photometric data}\label{phot_data}

We performed broad-band photometry using SCORPIO-1 data, obtained in the $U$, $B$, $V$, $R_C$, and $I_C$ filters immediately before the spectral observations. These measurements were used to derive reliable estimates of the total luminosity and interstellar extinction toward Var 2 via model fitting. The total exposure times of direct images were 60 s ($U$), 50 s ($B$), 30 s ($V$), 50 s ($R_C$), 30 s ($I_C$). Pre-proccessing of raw data and frame stacking were performed in the \textsc{midas} environment.
 
\begin{figure*}
\centering
\includegraphics[scale=0.4, angle=0]{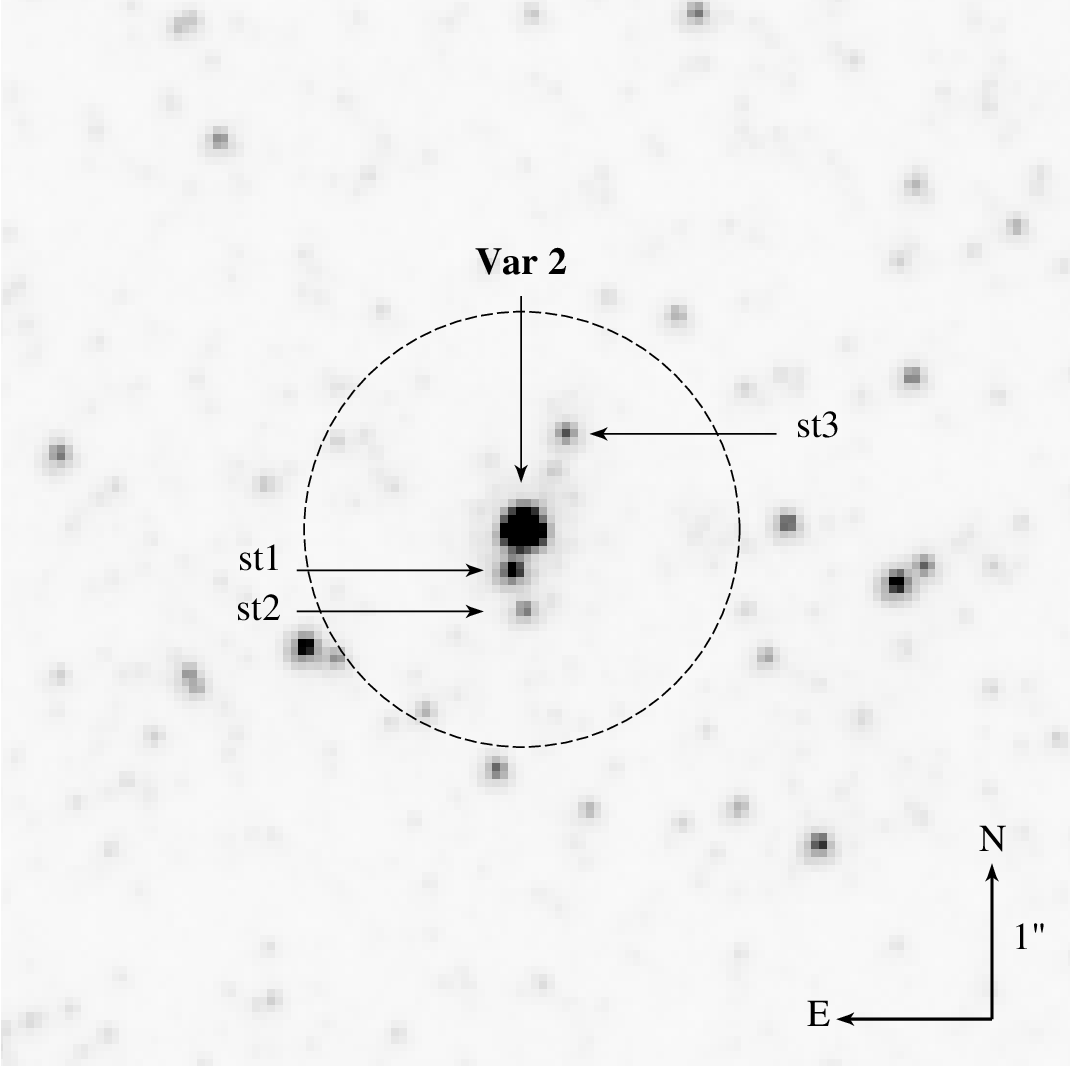}
\caption{Image of the Var\,2 vicinity in the $F814W$ filter obtained with ACS/WFC camera on HST. The 1.4\arcsec{} radius aperture used for the ground-based photometry marked with dashed circle. In addition to Var\,2, we marked the three brightest nearby sources (st1, st2, and st3), whose flux was accounted for in the ground-based photometry.}
\label{im:hst}
\end{figure*}

The spatial resolution of ground-based telescopes does not allow to resolve nearby faint sources in the vicinity of Var\,2. Figure~\ref{im:hst} shows Hubble Space Telescope (HST) images with the brightest stars marked, which enter the aperture during photometry of direct images from ground-based telescopes. Total flux from Var\,2 and nearby sources was measured in 1.4\arcsec{} radius aperture using \textsc{apphot} package in the \textsc{iraf} environment. The local background level was estimated within an annulus with inner and outer radii of 2.2\arcsec{} and 3.6\arcsec{}, respectively. The absolute calibration of the obtained stellar magnitudes was based on the measurements of selected reference stars from the \citet{Massey2007} catalog. We found the total stellar magnitudes $U_{sum}=17.37\pm0.07\,$mag, $B_{sum}=18.15\pm0.04\,$mag, $V_{sum}=18.20\pm0.04\,$mag, $R_{C,sum}=18.10\pm0.04\,$mag and $I_{C, sum}=17.93\pm0.05\,$mag.
 
The contribution of the sources adjacent to Var\,2 to the total flux was accounted for using archival Hubble Space Telescope (HST) images obtained with the WFC3 camera in the $F336W$ filter, and the ACS/WFC camera in the $F475W$, $F625W$, and $F814W$ filters. We corrected the total flux only for the contributions from st1, st2 and st3, as the influence of remaining sources within the 1.4\arcsec{}{} aperture is negligible, being comparable to the measurement errors. Their fluxes were measured using the \textsc{apphot} package with aperture corrections derived from encircled energy tables\footnote{Encircled energy tables are available at https://www.stsci.edu/}. The derived magnitudes were converted to the $U$, $B$, $V$, $R_C$, and $I_C$ filters using the \textsc{pysynphot} package, using \citet{Castelli2003} atmosphere models that best fitted the observed spectral energy distributions. The final stellar magnitudes of Var\,2 were determined to be $U=17.61\pm0.09\,$mag, $B=18.34\pm0.05\,$mag, $V=18.40\pm0.05\,$mag, $R_C=18.29\pm0.05\,$mag and $I_C=18.10\pm0.09\,$mag. The continuum in the optical spectrum of Var\,2 was additionally corrected for the contribution from neighboring stars.
 
The derived apparent magnitudes agree within $0.1-0.2$\,mag with the results of numerous ground-based photometric studies of Var\,2 in the M33 galaxy conducted over the past half-century \citep{Humphreys1975, Humphreys1978, Kurtev1999, Massey2007, Massey2016, Martin2017}. According to measurements from the Zwicky Transient Facility (ZTF) monitoring program \citep{Bellm2019, Masci2019}, the mid-term variability of Var\,2 is at the level of 0.2 mag, which is typical for "dormant" LBVs \citep{Maryeva2022} and some supergiants \citep{Guzik2024}. Therefore, it can be concluded that Var\,2 has not exhibited significant variability on various timescales.

\section{Modeling of the extened atmosphere}
\label{modeling_section}

The fundamental parameters of Var~2 were derived from the obtained spectral and photometric data using spherically symmetric non-LTE models of extended atmospheres calculated using the {\tt CMFGEN} \citep{HillierMiller1998} code. Due to the high complexity of such calculations, we focused on the determination of effective temperature and mass-loss rate during modeling, while other important model parameters were fixed at the initial stage of the work to simplify the selection of the optimal model.

\begin{figure*}[t]
     \centering
         \includegraphics[width=\linewidth]{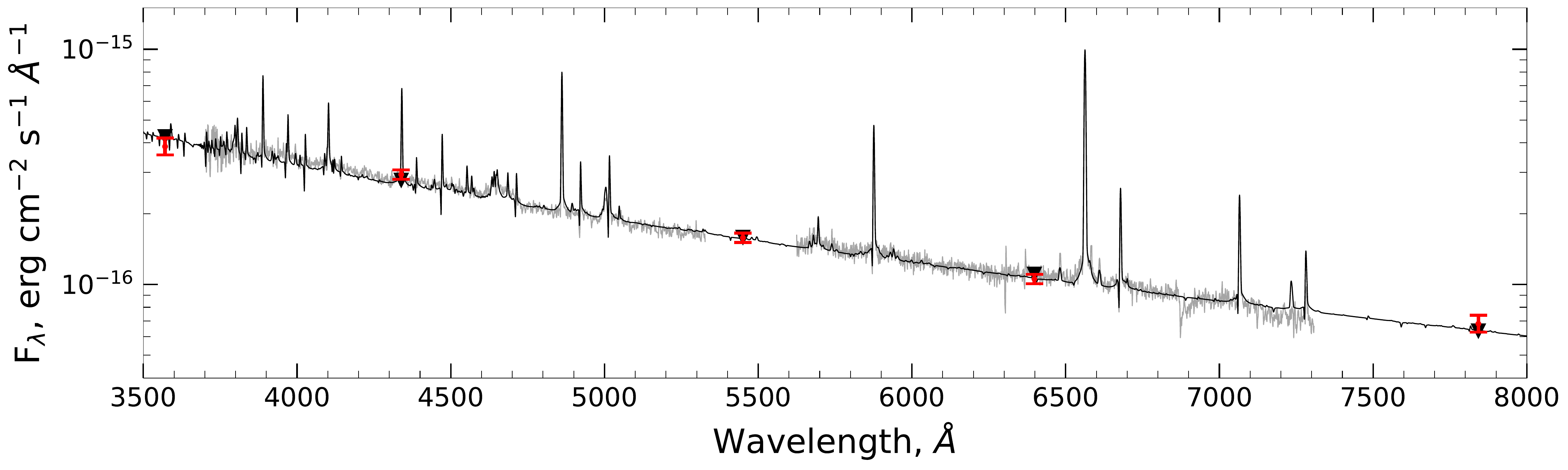}
        \caption{The optical spectrum (grey line) and best-fit baseline model (black line) of Var\,2 in the absolute units. Observed and model photometry are shown by red bars and black triangles, respectively.}
        \label{fig:sed}
\end{figure*}

The effective temperature of $T_{\rm eff}=24.4\,$kK was estimated from the relative balance of the helium lines: He\,II~$\lambda$4686, the He\,I~$\lambda$6678 singlet, and the He\,I~$\lambda$5876, $\lambda$7065 triplets. By varying the mass-loss rate $\dot{M}$, we had found the optimal match between the model and observed profiles of the strongest hydrogen lines of the Balmer series at $\dot{M}=2.1\times10^{-5}\,M_{\odot}\,\text{yr}^{-1}$. The luminosity $L=6.5\times10^{5}\,L_\odot$ had been determined by fitting the photometric data presented in Section~\ref{phot_data} adopting the distance $d=920\,$kpc to the galaxy M33 \citep{Jacobs2009} and using the absorption curves from \cite{Calzetti1994}. We have derived color excess value of $E(B-V)=0.16\pm0.01$, which is in good agreement with the average value of $E(B-V)\approx0.13$ for OB stars in the galaxy M33 \citep{Massey1995, Massey2007}. The observed optical spectrum and best-fit baseline model of Var\,2 in the absolute units, corrected for distance and reddening, along with photometry are presented in Figure~\ref{fig:sed}.

The of wind velocity was parametrized by the isothermal scale height at the wind base, which turns into a simple exponential $\beta$-law in the upper layers of the atmosphere \citep{Najarro1997, Hillier2003}:

\begin{equation}\label{betalaw}
    v(r)=\frac{v_0 + (v_{\infty} - v_0)(1 - R_* / r)^\beta}{1 + (v_0/v_\text{core})\exp[{(R_* - r)/h_\text{eff}}]},
\end{equation}

where $\beta$ is the exponent describing the wind acceleration with distance from the center of a star; $v_\text{core}$ is the velocity at the inner hydrostatic radius $R_*$; $h_\text{eff}$ is the effective isothermal scale of the atmosphere; $v_0$ is the parameter determining the gas velocity distribution in the transition zone between the lower layers of the atmosphere and the wind; $v_\infty$ is the terminal wind velocity.

The value of $v_\text{core}$ was chosen such that the optical depth at the hydrostatic radius $R_*$ remained within $50 \leqslant \tau_\text{Ross} \leqslant 200$ during modeling. The value of $v_{0}=20$\,km\,s$^{-1}$ corresponded to the sound speed $v_\text{sonic}\approx18$\,km\,s$^{-1}$ near the wind transit zone. The choice of $h_\text{eff}=0.1\,R_*$ was based on the relation between the gas and radiation pressure to gravity in the inner parts of the wind. We note that hydrostatic structure of the atmosphere cannot be exactly described with the present approximation.

The results of numerous studies of LBVs do not allow us to make unambiguous conclusions about the dynamical properties of their wind. There are several examples of sucessfull applications of $\beta\sim1-3$ \citep{Najarro2009, Groh2009, Kostenkov2020, Maryeva2022} as an analytic form of wind velocity law. The results of radiation-hydrodynamic modeling of the wind of such stars also differ significantly \citep{Vink2018, BerniniPeron2025}. To reduce the number of free parameters of our models we had chosen $\beta=1.0$ as an initial approximation of the Var\,2 wind model. This value is typical for the winds of the hottest blue stars \citep{Puls1996, Muijres2012, GormazMatamala2021}.

The absorption components of the Balmer series hydrogen and He\,I line profiles constrain the terminal wind velocity of Var\,2 as $v_\infty\approx230\,$km\,s$^{-1}$ \citep{Humphreys2014}. However, such terminal velocity overestimates the widths of the emission components of the lines in the models relative to observations. As the modeling results shows, the general appearance of the calculated line profiles is closer to the observed ones for a slightly lower value of the terminal velocity $v_\infty=210\,$km\,s$^{-1}$.

Observations in a wide range of wavelengths indirectly indicate on a significant heterogeneity of matter in the winds of hot massive stars \citep{Nugis1998, Hawcroft2021, Rubke2023}. In our calculated wind models, the presence of optically thin clumps of matter separated by empty space was taken into account \citep{Hillier1999}. This approach involves adjusting the gas density using the volume filling factor $f$, defined as the ratio between the uniform (smoothed) density to the density of clumps at a given radius. The distribution of $f(r)$ in the wind was parameterized using the expression from \cite{Najarro2009}:

\begin{equation}\label{clump_param}
f(r)=f_1 + (1 - f_1)\exp\left(\frac{-v(r)}{f_2}\right) + (1 - f_1)\exp\left(\frac{v(r) - v_\infty}{f_3}\right)
\end{equation} 

where $f_1$ is the maximum value of the volume filling factor, and $f_2$ and $f_3$ define the velocities at which the degree of wind inhomogeneity increases and decreases, respectively. This distribution is in good agreement with observations \citep{Puls2006, Najarro2011} and theoretical models of stellar winds \citep{Runacres2002, Sundqvist2013}.

The value of $f_1$ is determined by comparing the intensities of emission lines that depend on the density linearly (resonance lines in the ultraviolet) and quadratically (recombination lines in the optical or infrared ranges) \citep{Prinja2013, Verhamme2024}. The degree of wind inhomogeneity can also be estimated from the relative intensity of the electron scattering wings of the strongest recombination lines, for example H$_\alpha$ \citep{Hillier1991}. Typically, the $f_1$ value for winds of hot stars lies in the range of $f_1 \approx 0.02-0.1$ \citep{Puls2008, Krticka2017, Bouret2021}. However, higher values of $f_1 \approx 0.15-0.5$ were obtained for some LBV/cLBV \citep{Najarro2001, Maryeva2019, Kostenkov2020}. Preliminary modeling results showed that the wings of the Balmer series hydrogen lines and some He\,I lines in the Var\,2 spectrum are best described by $f_1\approx0.1$.

We adopt the following density structure for our models. The wind heterogeneity begins to increase as the wind velocity approaches the sound speed, reaching a maximum above the critical point. After passing the plateau corresponding to the minimum values of $f_1$, the density of clumps decreases, and upon reaching the terminal velocity $v_\infty$ the ejected matter again becomes homogeneous. Thus, the value of $f_2$ was taken equal to the wind speed near the transition zone of the extended atmosphere $f_2=v_0=20$\,km\,s$^{-1}$. Independent of $f_1$ and $f_2$ determination of the value of $f_3$ is possible only using data in the infrared and radio data \citep{Najarro2009, Najarro2011, Puls2006}. For our models we assumed that the wind clumping peaks at $r\approx2\,R_*$ in accordance with the results of hydrodynamic modeling of the wind of the prototype of an early O-supergiant presented in \cite{Sundqvist2013}. A similar distribution of clumps in the wind was empirically obtained for the O4\,I(n)f star $\zeta\,$Pup \citep{Puls2006, Najarro2011}. At this distance, the wind in the model accelerates to $v(r)\approx0.5\,v_\infty$, therefore, to obtain the minimum $f_1$ at $r\approx2\,R_*$, the value of $f_3$ was taken equal to $f_2$. An analogous parameterization of the distribution of inhomogeneities was also used during the modeling of the galactic cLBV MN112 \citep{Maryeva2022} which is somewhat cooler (spectral type B1-2\, Ia-0ep).

The hypothesis that clumps begin to form deep in the extended atmosphere near sound speed is supported by the results of modeling of the winds of early-type stars \citep{Bouret2003, Bouret2005, Hillier2003}. However, in some cases significantly higher values of the velocity at which the wind becomes inhomogeneous were adopted for an optimal description of the observed spectrum \citep{Bouret2008, Sundqvist2011, Kostenkov2020}. In addition, it is often assumed that the wind remains inhomogeneous even in the outer parts (\citet{BerniniPeron2024, Gvaramadze2018, Kostenkov2020}, also see the references above). In such cases a simpler form of expression (1) with $f_3=0$ \citep{Hillier1999} is used. For example, in both papers \cite{Kostenkov2020} and \cite{Maryeva2022}, devoted to the study of cLBV MN112, a good agreement between the model and observed spectral line profiles in the near infrared range was obtained with different values of $\beta$ in the velocity law and different distributions of clumps in the wind.

To study the influence of the parameterization of the volume filling factor $f$ at a fixed value $\beta=1.0$ on the obtained model spectra, we had constructed two baseline models with different distributions of wind inhomogeneities: $f_1=0.1$, $f_2=f_3=20$\,km\,s$^{-1}$ (model~A) and $f_1=0.5$, $f_2=20$\,km\,s$^{-1}$, $f_3=0$ (model~B). The value of $f_1$ in model~B was chosen such that the electron scattering wings of the Balmer hydrogen lines were identical to those in model~A. The temperature and mass-loss rate in model~B were also slightly corrected (within 10\%) in accordance with the ratio of the equivalent widths of the H\,I and He\,I lines in the reference model~A, the other wind parameters remained unchanged.

\begin{figure*}[t]
     \centering
         \includegraphics[width=0.70\linewidth]{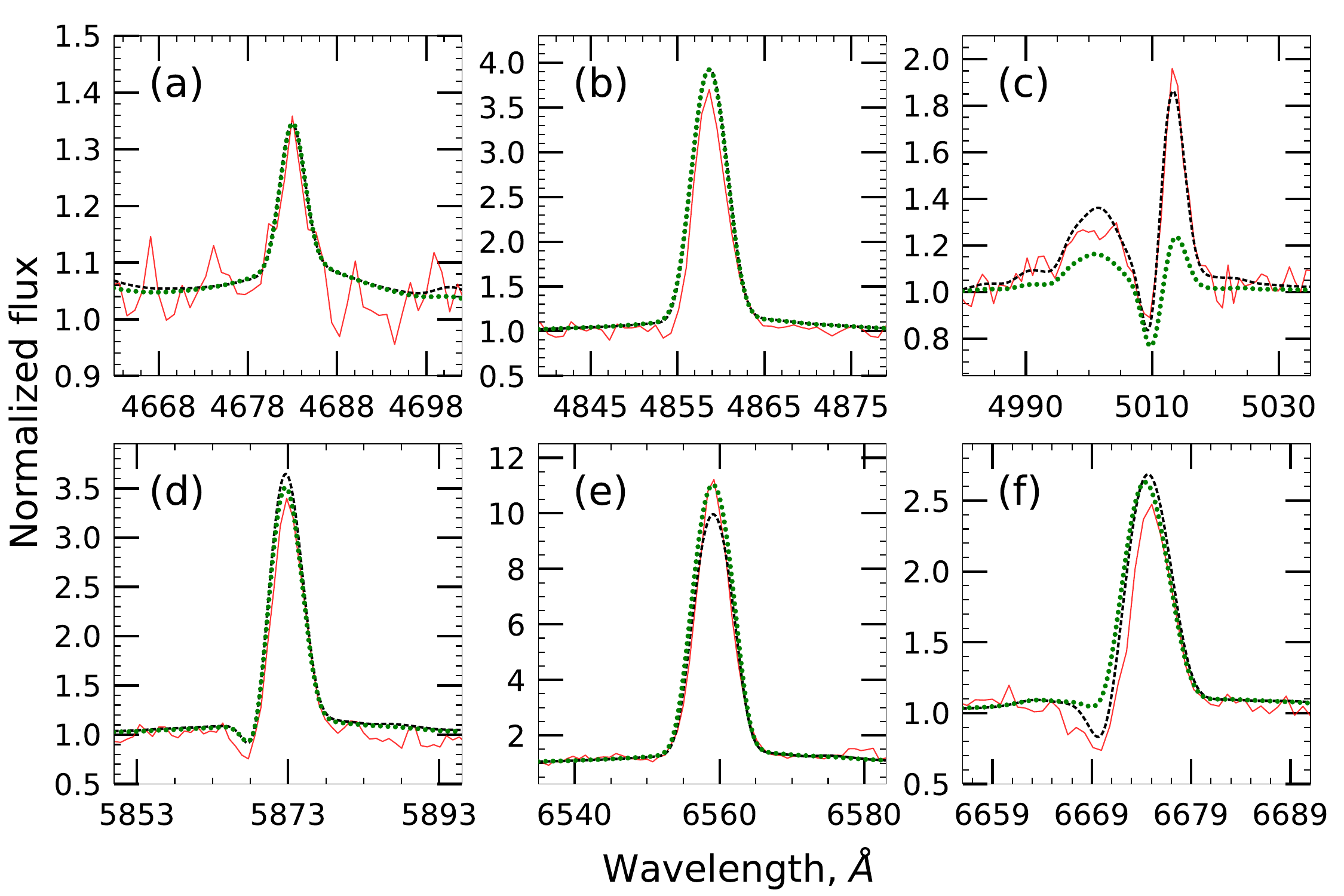}
        \caption{Comparison of the line profiles for several lines in models with different distributions of wind inhomogeneties: (a) He\,II~$\lambda4686$; (b) H$_\beta$; (c) N\,II~$\lambda\lambda$4994--5007, He\,I~$\lambda5015$; (d) He\,I~$\lambda5876$; (e) H$_\alpha$ (f) He\,I~$\lambda$6678. The black dashed and green dotted lines correspond to models A~and~B (see text). Normalized spectrum of Var\,2 is shown with red solid line.}
     \label{fig:clump_comp}
\end{figure*}

As can be seen in Figure~\ref{fig:clump_comp}, model~A and model~B demonstrate\footnote{Model spectra were shifted by 170\,km\,s$^{-1}$ to the short-wave region for clarity.} similar line profiles of the Balmer series hydrogen, ionized helium He\,II~$\lambda$4686 and the triplets of neutral helium He\,I~$\lambda5876$, $\lambda$7065. At the same time, the He\,I~$\lambda$5017, $\lambda$6678 singlets differ significantly in the presented models: in model~B the He\,I~$\lambda$5017 line emission is markedly weaker, and the He\,I~$\lambda$6678 line profile almost lacks an absorption component. 

Such a discrepancy in the He\,I singlets between the models is explained by the dramatic difference between optical depth of the He\,I continuum related to the change in the density distribution in the wind between models. The upper level of 5017\,\AA{} (1s\,2s\,$^1$S--1s\,3p\,$^1$P$^{\rm o}$) and lower level of 6678\,\AA{} (1s\,2p\,$^1$P$^{\rm o}$--1s\,3d\,$^1$D) transitions are fed directly from the ground state by the 537\,\AA{} (1s$^2$\,$^1$S--1s\,3p\,$^1$P$^{\rm o}$) and 584\,\AA{} (1s$^2$\,$^1$S--1s\,2p\,$^1$P$^{\rm o}$) resonance transitions, respectively. Efficient pumping through these channels occurs only in the model\,A with high optical depth of the He\,I continuum, where a sufficient density in the continuum-forming region maintains a substantial ground-state population. The reduction in density within these critical layers significantly lowers the ground-state population, suppressing both pumping routes and weakening the two spectral features together in the model\,B. Furthermore, escaped photons alter the ionization state of iron in the outer parts of the wind, increasing number of Fe\,IV ions. It has a notable effect on the radiation field near the He\,I~$\lambda$584 resonance line \citep{Najarro2006}. As shown in \cite{Najarro2006}, the loss of photons in the Fe\,IV lines near 584\,\AA{} significantly changes the population of the 1s\,2p\,$^1$P$^{\rm o}$ level. As a consequence, named transition affects the profile and intensity of the He\,I lines associated with it. As shown in Fig.~\ref{fig:clump_comp}, model~A fits the observed He\,I~$\lambda5017$, $\lambda6678$ lines profile better in comparison with model~B that uses simpler form of the volume-filling factor $f(r)$ distribution in the wind.

Strong emission in important diagnostic lines, caused by the dense wind, significantly distorts their absorption components, which, together with the moderate signal-to-noise ratio and low resolution of our spectrum, prevents a reliable independent determination of the stellar rotation and turbulent velocity. Overall, spectrum of Var\,2 does not show any notable signs of rapid rotation of the star, similar to "dormant"{} LBV P\,Cygni \citep{Najarro1997}, which is assumed to be a relatively slow rotator with $v_{\rm rot}/v_{\rm crit}\sim0.2-0.3$ \citep{Groh2009_hrcar}. The turbulent velocity $v_{\rm turb}=15$\,km\,s$^{-1}$ was fixed at the upper limit of the range of typical values $v_{\rm turb}=10-15$\,km\,s$^{-1}$ found for early B \citep{Trundle2004, Crowther2006, Markova2008} and late O \citep{Smith1998, Crowther2002} supergiants. This value correspond to the sound speed in the wind transition region and was found to be slightly better in terms of describing the observed depth of the absorption component of some He\,I line profiles than lower $v_{\rm turb}$ values.

We had derived the hydrogen abundance ${\rm X}_{\rm H}\approx43\%$ in the atmosphere of Var\,2 with the ratio between the equivalent widths of the Balmer series lines and the triplet lines of neutral helium He\,I~$\lambda5876$, $\lambda7065$. The mass fraction of carbon ${\rm X}_{\rm C}=0.17\,{\rm X}_\odot$ and nitrogen ${\rm X}_{\rm N}=1.70\,{\rm X}_\odot$ had been determined with C\,III~$\lambda5696$ and N\,II~$\lambda\lambda$4994--5010 lines, respectively. To estimate the abundances of other chemical elements, we used value of the average metallicity ${\rm Z}\approx0.5\,{\rm Z}_\odot$ in the galaxy M33 \citep{Kang2012}. Due to the absence of bright and distinct oxygen lines in the optical spectrum, its abundance was assumed to be ${\rm X}_{\rm O}=0.03\,{\rm X}_\odot$ according to evolutionary models with ${\rm Z}=0.006$ \citep{Eggenberger2021}. The following ions were included in the Var\,2 wind models: H\,I, He\,I-II, C\,II-IV, N\,II-IV, O\,II-V, Ne\,II-IV, Mg\,II-III, Si\,II-IV, P\,III-IV, S\,III-V, Cl\,IV-VI, Ar\,III-V, Ca\,III-VI, Cr\,III-VI, Mn\,III-VI, Fe\,III-VI, Nk\,III-VI, Co\,III-VI. The total number of bound-bound transitions in the calculated models was approximately $\approx1.3\times10^{5}$.

Due to the presence of dense wind obscuring the lower layers of the atmosphere, determining the mass of the star from the diagnostics of absorption lines is difficult. Estimates of the possible mass of Var\,2 were obtained from the energy balance equation for the stellar wind \citep{Petrov2016}:

\begin{equation}\label{wind_energy_balance_eq}
\dot{M}\left(\frac{v_\infty^2}{2} + \frac{GM_*}{R_*}\right)=\dot{M}\int_{R_*}^{\infty}\left(- \frac{1}{\rho}\frac{{\rm d}P}{{\rm d}r} + g_{\rm rad}\right){\rm d}r,
\end{equation}
where $g_{\rm rad}$ is the wind acceleration created by radiative pressure. Thus, for $v_\infty=210\,$km\,s$^{-1}$ the optimal mass of the star is $M_*\approx24\,M_\odot$.

\begin{figure*}[t]
     \centering
         \includegraphics[width=\linewidth]{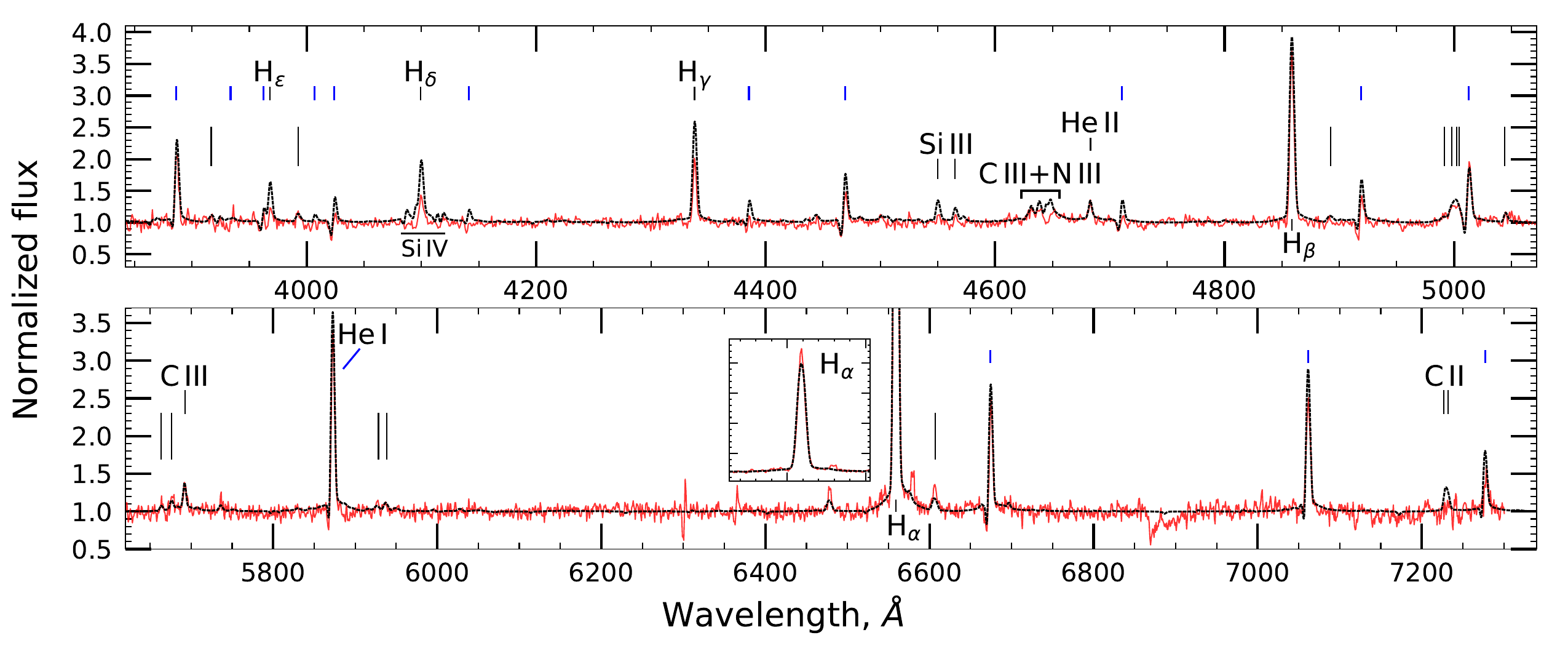}
        \caption{The observed normalized spectrum (red solid line) and the best-fit baseline model (black dashed line) of Var\,2. The He\,I and N\,II lines are indicated by short blue and long black dashes, respectively.}
        \label{fig:model_norm}
\end{figure*}

\begin{table}
\centering
\caption {Key parameters of the two extended atmosphere models of Var\,2: an analytical velocity law model and a self-consistent hydrodynamic model. The remaining wind characteristics of the presented models are discussed in detail in Sections~\ref{modeling_section} and \ref{hydro_section}. $R_{2/3}$ and $T_{\rm eff}$ are radius and temperature at $\tau_{\rm Ross}=2/3$, $T_*$ is the temperature at hydrostatic radius $R_*$ ($\tau_{\rm Ross} \approx 150$).}
\label{tab:par}
\medskip
\begin{tabular}{l|c|c}
\hline
Parameter & Base model & Hydro model  \\
\hline
$L$, [$L_\odot$] & $6.5\times10^{5}$ & $6.5\times10^{5}$ \\
$M_*$, [$M_\odot$] & 24  & 23 \\
$\dot{M}$, $[M_{\odot}\,\text{yr}^{-1}]$ & $2.1\times10^{-5}$ & $2.1\times10^{-5}$  \\
$T_{\rm eff}$, [kK] & $24.4$ & $24.8$  \\
$R_{2/3}$, [$R_\odot$] & $45.0$ & $43.7$  \\
$T_{*}$, [kK] & $34.2$ & $32.3$  \\
$R_{*}$, [$R_\odot$] & $23.0$ & $25.7$  \\
$\beta$ & $1.0$ & --- \\
$v_{\infty}$, [km\,s$^{-1}$] & 210 & 230  \\
$f_1$ & $0.1$ & $0.1$ \\
$v_{\rm turb}$, [km\,s$^{-1}$] & 15 & 15 \\ 
${\rm X}_{\rm H}$, [\%] & 43 & 43 \\ 
\hline
\end{tabular}
\end{table}

The observed normalized spectrum and the best-fit model are shown in Figure~\ref{fig:model_norm}. The resulting model shows good agreement between the model and the observed brightest hydrogen lines of the Balmer series, He\,I, N\,II, and C\,III. However, there are features of the spectrum that could not be reproduced exactly, for example, the Balmer decrement, depth of the absorption component of some He\,I lines or the silicon lines Si\,III/IV. This discrepancy between the observations and the model can probably be due to the complex structure of the distribution of inhomogeneities in the wind (for example, see the H$_\alpha$/H$_\beta$ ratio in Figure~\ref{fig:clump_comp}) which can be optically thick. This effect can significantly affect the observed line profiles \citep{Prinja2010, Petrov2014}.

The main parameters of the presented extended atmosphere model are listed in Table~\ref{tab:par}. It is important to note that the properties of the obtained model strongly depend on the assumptions regarding wind parameters described in this section. Reliable determination of these wind parameters generally requires multi‑wavelength spectral data: for instance, UV spectra are typically needed to determine the mass-loss rate independently of clumping or to estimate terminal velocity \citep{Groh2009, Bouret2012, Rickard2022}, while IR data are essential for constraining the velocity profile \citep{Najarro1997_iso_sws, Kostenkov2020} and clumps stratification \citep{Najarro2009, Najarro2011}. Because no spectra of Var\,2 are available outside the optical range, we adopted constraints on these parameters from studies of other massive stars, focusing our modeling on the mass-loss rate and temperature, while keeping other parameters fixed. The moderate signal-to-noise ratio and low resolution of our spectrum do not significantly affect determination of these parameters due to high intensity and sufficient width of hydrogen and helium lines in the optical range.

\section{Wind dynamics}\label{hydro_section}

At the next stage of the Var\,2 study, we had constructed a self-consistent wind model. The model was designed to take into account the balance of radial forces in the extended atmosphere of the star. A correct description of the wind velocity law allowed us to study the outflow features of this LBV in more detail and to refine some of the model parameters.

The continuity equation for a spherically symmetric wind is:

\begin{equation}\label{cont_eq}
\dot{M}=4\pi r^2 v(r) \rho(r) f(r),
\end{equation} 
while the equation of momentum conservation can be written as

\begin{equation}\label{mom_eq}
v\frac{dv}{dr}= - \frac{1}{\rho}\frac{{\rm d}P}{{\rm d}r} - \frac{GM_*}{r^2} + g_{\rm rad}.
\end{equation}

We adopted constant sound speed in the wind $a={\rm const}$ for our models. \cite{Sander2017} showed that contribution of thermal expansion to wind acceleration at $v>a$ quickly becomes insignificant compared to the radiation pressure. For the same reason, to simplify the calculations, the gas had been considered homogeneous while calculating the thermal pressure. However, we note that a significant change in the population of levels due to an increased level of recombinations at a higher density in clumps and, as a consequence, an increase in the radiation pressure was taken into account in the calculated model. By combining equations \ref{cont_eq} and \ref{mom_eq}, in the case of an ideal gas ($P=a^2 \rho$), the dynamics of an isothermal homogeneous wind can be described by the equation

\begin{equation}\label{eom}
\left(v - \frac{a^2}{v}\right)\frac{{\rm d}v}{{\rm d}r}=\frac{2a^2}{r} - \frac{GM_*}{r^2} + g_{\rm rad}
\end{equation}
with a singularity at the critical point $v(r_{\rm c})=a$. The velocity gradient in the critical point determines the nature of the solutions of this equation. In case of a monotonically accelerating wind, the condition $\frac{{\rm d}v}{{\rm d}r}>0$ must be satisfied at $v(r)=a$ \citep{Parker1964}. The critical point $r_{\rm c}$ can be found from the equation:

\begin{equation}\label{crit_point_eq}
2a^2r_{\rm c} - GM_*  + g_{\rm rad}(r_{\rm c})r_{\rm c}^2=0.
\end{equation}
Next, the wind acceleration at $r=r_{\rm c}$ is calculated using de l'H\^{o}pital's rule \citep{LamersCassinelli1999}:

\begin{equation}\label{lhop_crit_point}
\left(\frac{{\rm d}v}{{\rm d}r}\right)_{r_{\rm c}}=\sqrt{-\frac{a^2}{r_{\rm c}^2} + \frac{GM_*}{r_{\rm c}^3} + \frac{1}{2}\left(\frac{{\rm d}g_{\rm rad}}{{\rm d}r}\right)_{r_{\rm c}}}.
\end{equation}
The smooth transition between the subsonic and transonic wind zones suggests that

\begin{equation}
\lim_{r \to r_{\rm c}-}v(r) = \lim_{r \to r_{\rm c}+}v(r)
\end{equation}
and

\begin{equation}
\lim_{r \to r_{\rm c}}\left.{\frac{{\rm d}v(r)}{{\rm d}r}}\right\vert_{r < r_{\rm c}} = \lim_{r \to r_{\rm c}}\left.{\frac{{\rm d}v(r)}{{\rm d}r}}\right\vert_{r > r_{\rm c}}.
\end{equation}

Thus, the complete form of a velocity distribution in the stellar wind can be found by numerically integrating the equation \ref{eom} up and down the radius from the critical point using the acceleration value obtained by the formula \ref{lhop_crit_point}. In this paper, the solution of the equation \ref{eom} was carried out by the explicit Runge-Kutta method of 5(4) order with an adaptive step size using the Dormand-Prince formulas \citep{DormandPrince1980}. The new wind velocity structure obtained by solving the equation \ref{eom} is not consistent with the old radiation field, so it is further used to calculate the new atmospheric model using the {\tt CMFGEN} code. As studies by various groups of authors have shown, the optimal way to achieve stable convergence of such iterative calculations is to correct the model parameters based on the balance of forces at the critical point \citep{Sander2017, Sundqvist2019}.

Unlike the other works, our calculations are based on the variation of the stellar mass instead of the mass-loss rate $\dot{M}$ or the critical point radius $r_{\rm c}$ to facilitate more straightforward model fitting and comparison. The new value of the stellar mass at each stage was determined using the expression \ref{crit_point_eq} for a fixed value of $r_{\rm c}$ chosen at the initial stage of the calculations. These iterations are repeated until the convergence condition is reached \citep{Sundqvist2019}:

\begin{equation}
\max\left\vert 1  - \frac{\left(v^{\rm old} - \frac{a^2}{v^{\rm old}}\right)\frac{{\rm d}v^{\rm old}}{{\rm d}r} - \frac{2a^2}{r} + \frac{GM^{\rm old}_*}{r^2}}{g^{\rm new}_{\rm rad}} \right\vert < 1 \times 10^{-2}
\end{equation}
with the stellar mass correction not larger than 1\% between a new model and a previous one. Improving convergence stability, we applied half of the velocity correction at each grid point in successive iterations. To minimize numerical artifacts, wind points beyond $r\gtrsim 100\,r_{\rm c}$ were excluded from the uncertainties calculation.

The critical point radius $r_{\rm c}$ choice in these self-consistent wind models determines the ionization state of matter in the extended atmosphere. We had analyzed synthetic spectra of the models with different $r_{\rm c}$ calculated on the basis of the best extended atmosphere model Var\,2, presented in Section~\ref{modeling_section}. The study of these spectra shows that the helium line balance closest to observations is achieved at $r_{\rm c}\approx2.4\times10^{12}\,$cm. This value turned out to be quite close to the radius of the sonic point $r_{\rm c}\approx2.3\times10^{12}\,$cm in the baseline model. At $r_{\rm c}\lesssim2.4\times10^{12}\,$cm the optical depth of the He\,I continuum markedly drops, which distorts the profiles of He\,I singlet lines (for example, see Figure~\ref{fig:clump_comp}). On the other hand, the ionized helium line He\,II~$\lambda$4686 turned out to be notably underestimated relative to observations in models with $r_{\rm c}\gtrsim2.4\times10^{12}\,$cm. In all hydrodynamic wind models computed in this work, the mass-loss rate remained constant and matched the value $\dot{M}=2.1\times10^{-5}\,M_{\odot}\,\text{yr}^{-1}$ derived in Section~\ref{modeling_section}.

\begin{figure*}[t]
     \centering
         \includegraphics[width=0.6\linewidth]{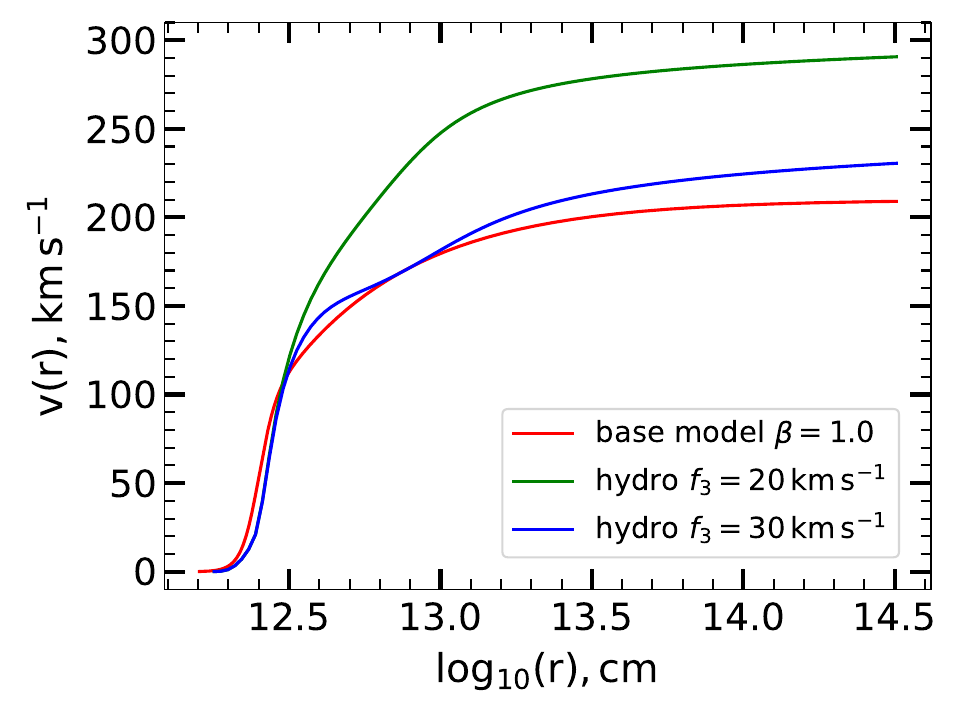}
        \caption{Wind velocity profiles for the baseline extended atmosphere model of Var\,2 (red solid line) presented in Section~\ref{modeling_section} and two self-consistent models calculated on its basis with different distributions of inhomogeneities in the wind. The lower boundary of the atmosphere for the self-consistent models corresponds to the Rossenland opacity $\tau_{\rm Ross} \approx 150$ similar to the initial baseline model.}
        \label{fig:hydro_vel}
\end{figure*}

\begin{figure*}[t]
     \centering
         \includegraphics[width=0.70\linewidth]{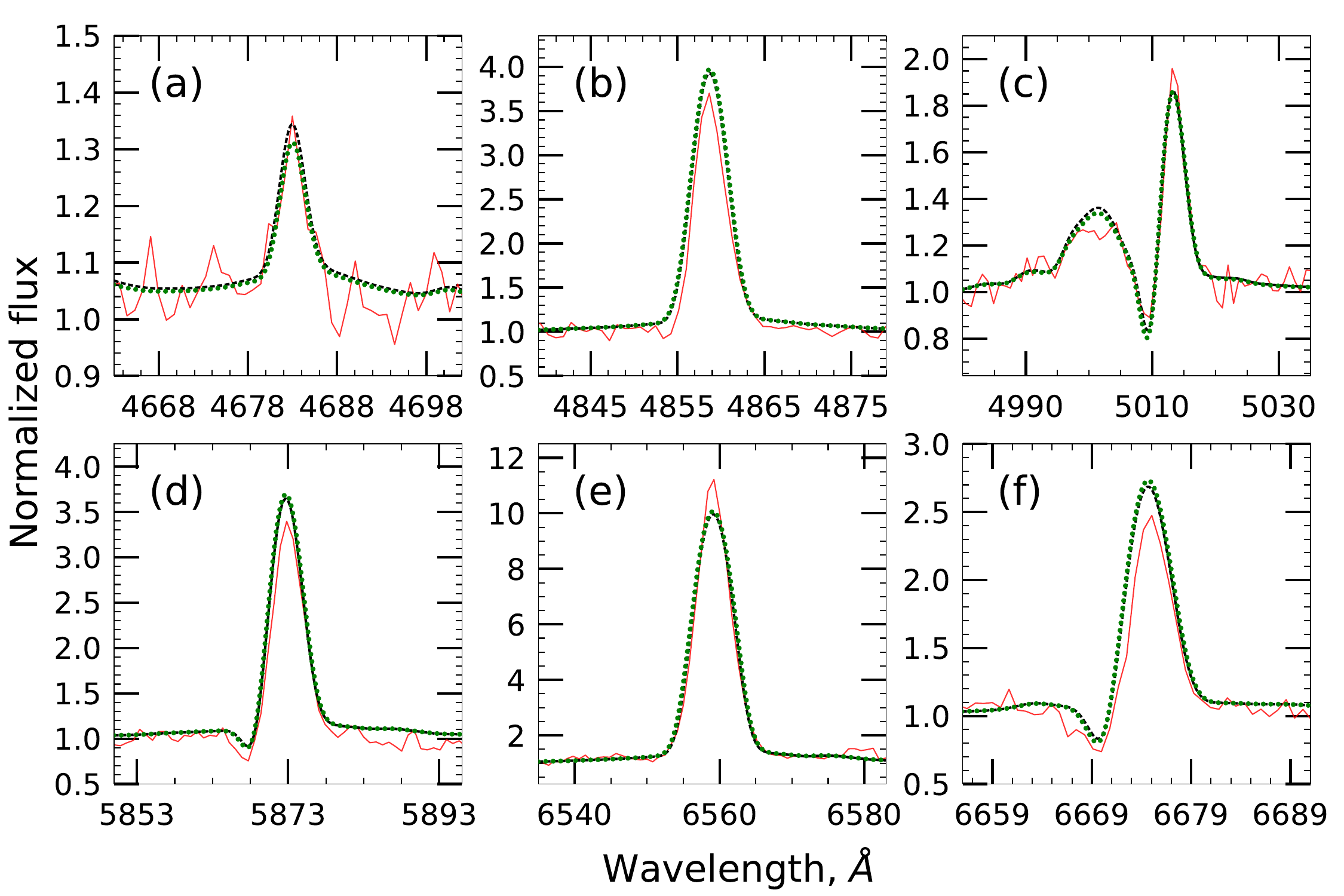}
        \caption{Comparison of the line profiles between models with analytical velocity distribution (black dashed) and self-consistent velocity structure (green dotted). The normalized observed spectrum of Var\,2 is shown with red solid line. The set of lines is the same as in Figure~\ref{clump_param}.}
     \label{fig:hydro_comp}
\end{figure*}

The velocity distribution in the self-consistent wind model with $r_{\rm c}=2.4\times10^{12}\,$cm is shown in Figure~\ref{fig:hydro_vel}. The terminal velocity of the wind in the resulting model $v_\infty\approx290$\,km\,s$^{-1}$ is significantly higher than the value $v_\infty\approx210$\,km\,s$^{-1}$ derived in Section~\ref{modeling_section}. At the same time, the mass value of $M_*\approx23\,M_\odot$ is slightly lower than reported above $M_*\approx24\,M_\odot$ for the baseline wind model. Taking into account the relation $v_{\rm esc}\propto v_\infty$ \citep{Lamers1995, Vink2018}, we can conclude that this discrepancy in the wind velocity is associated not only with the difference in the mass of the star in the models. We had calculated the terminal velocity of the hydrodynamic model using the expression \ref{wind_energy_balance_eq} with the variable sound speed and inhomogeneous density taken into account to determine the gas pressure. The obtained value of $v_\infty\approx310$\,km\,s$^{-1}$ is close to the terminal wind speed of the self-consistent model, so the difference in the obtained velocities is not related to the model calculation algorithm.

Such a discrepancy can be explained by the underestimation of wind acceleration in the initial model, which uses an analytical velocity law with $\beta=1.0$. As the wind velocity increases more rapidly above the critical point showing higher velocity gradient values, the efficiency of wind acceleration increases due to the growing contribution of optically thick lines to the total radiative acceleration, since $g^{\rm thick}_{\rm rad} \propto \frac{{\rm d}v}{{\rm d}r}/{\rho r^2}$.

Additional factor influencing the radiative acceleration is a dependence of the distribution of inhomogeneities in the wind on the velocity (see expression~\ref{clump_param}). Clumping enhances line-driven acceleration because it boosts recombination, thereby increasing the population of the atomic levels responsible for the driving transitions \citep{GormazMatamala2021}. An increased wind velocity at fixed clumping parametrization (with $f_1$, $f_2$, $f_3$ held constant) shifts the decline in clumps density to the outer wind, which boosts total line acceleration.

As shown by \cite{BerniniPeron2025}, the parameterization of inhomogeneities distribution in the wind in hydrodynamic calculations has a determining effect on the obtained relationships between the mass-loss rate and the terminal velocity. To find a model closer to the observations, we had calculated several self-consistent models with different distributions of clumps in the wind, varying $f_3$ to control radiative acceleration above the sonic point. The best agreement with the observed velocities was shown by the model with $f_1=0.1$, $f_2=20$\,km\,s$^{-1}$ and $f_3=30$\,km\,s$^{-1}$. Increasing the value of $f_3$ shifts the position of the minimum of the volume filling factor $f(r)$ to lower velocities. As a consequence, we obtain a reduced radiative pressure $g_{\rm rad}$ in the outer part of the wind without any significant change of the balance of forces near the critical point. The usage of this parametrization led to a lower terminal wind velocity $v_\infty=230$\,km\,s$^{-1}$ with almost unchanged stellar mass $M_* \approx23\,M_\odot$ compared to the initial self-consistent model with $f_3=20$\,km\,s$^{-1}$. The spectrum of the final hydrodynamic model with $f_3=30$\,km\,s$^{-1}$ is nearly identical to the one of baseline model presented in Section~\ref{modeling_section} (Figure~\ref{fig:hydro_comp}). The main parameters of the resulting self-consistent model are listed in Table~\ref{tab:par}. The luminosity $L=6.5\times10^5\,L_\odot$ in the hydrodynamic model agrees well with the observed photometry, as its wind velocity distribution with remain close to those of the baseline model that uses an analytical velocity profile with similar mass-loss rate.

\section{Discussion}

Through self-consistent non-LTE modeling of the optical spectrum of Var\,2, we obtained the main stellar and wind parameters of this LBV, including its current mass and wind velocity profile, which represents a significant advance in the study of such stars. The derived current mass, the luminosity and surface hydrogen abundance provide key constraints for testing and refining evolutionary models of massive stars. Here, we discuss the obtained results of modeling in the context of the possible evolutionary status of Var\,2. In the subsections below, we compare these stellar parameters with evolutionary tracks and match obtained wind properties with those of other massive stars. Next, we analyze the nearby stellar environment of Var\,2.

\subsection{Evolutionary status}

\begin{figure*}[t]
     \centering
         \includegraphics[width=0.49\linewidth]{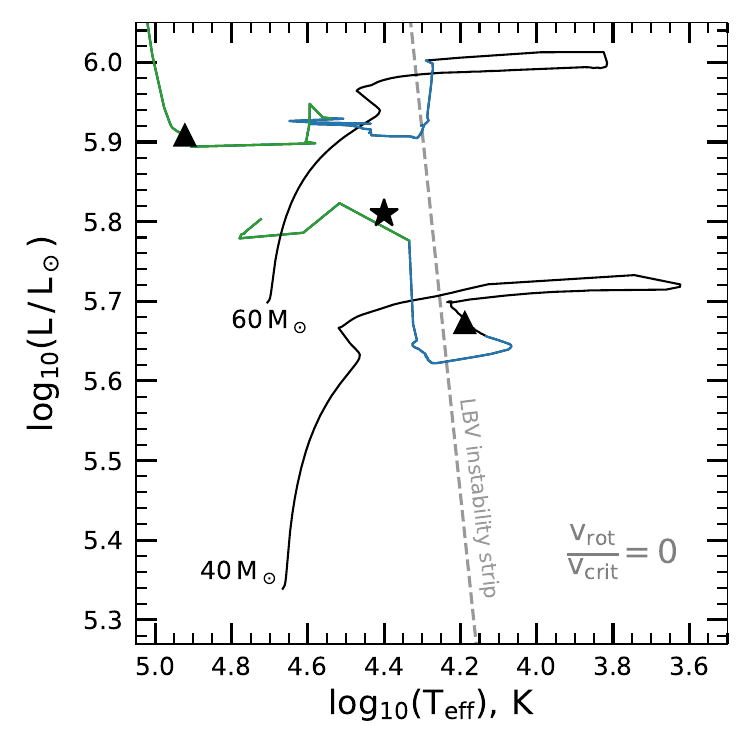}
         \includegraphics[width=0.49\linewidth]{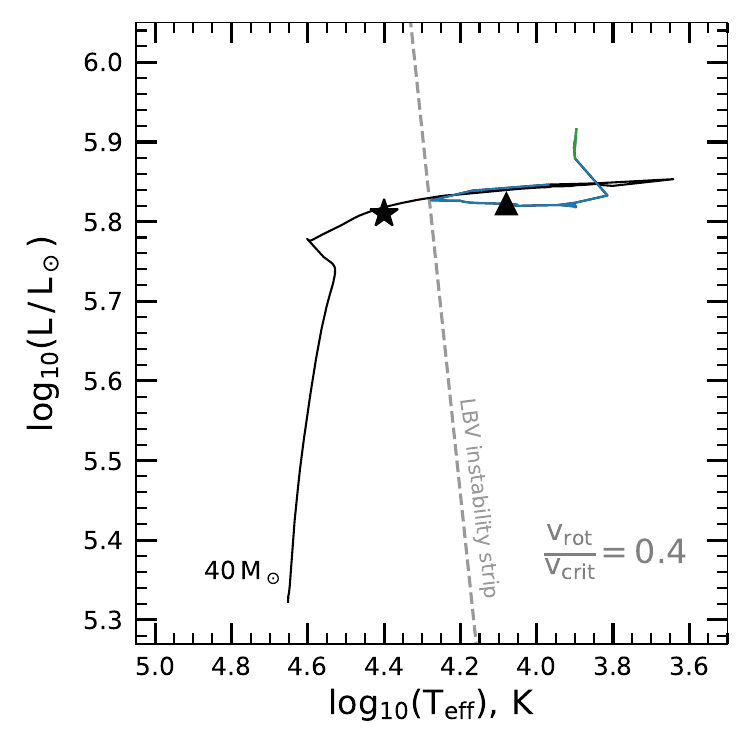}
        \caption{Position of Var\,2 (black star) on the Hertzsprung-Russell diagram along with the Geneva evolutionary tracks for the metallicity ${\rm Z}=0.006$  \citep{Eggenberger2021}, colored by the hydrogen abundance on the stellar surface: X$_{\rm H}>40\%$ (black solid line), $20\%<{\rm X}_{\rm H}<40\%$ (blue solid line) and X$_{\rm H}<20\%$ (green solid line). Left and right panels shows evolutionary tracks without rotation and with rotation $v_{\rm rot}/v_{\rm crit}=0.4$, respectively. Black triangle marks the position on track corresponding to the Var\,2 current mass $M\approx23\,M_\odot$. The LBV minimum instability strip is indicated by a grey dashed line with a slope taken from \cite{Groh2009_hrcar}}
     \label{fig:evol_tracks}
\end{figure*}

Figure~\ref{fig:evol_tracks} shows the location of Var\,2 on the Hertzsprung-Russell diagram for the luminosity $L=6.5\times10^5\,L_\odot$ and effective temperature $T_{\rm eff}=24.8\,$kK derived from the modeling. According to Geneva evolutionary models \citep{Eggenberger2021}, position of Var\,2 relative to non-rotating evolutionary tracks (Figure~\ref{fig:evol_tracks}, left panel) lies between tracks that correspond to the initial stellar masses of 40\,$M_\odot$ and 60\,$M_\odot$. These models predict that when a non-rotating star with $M_{\rm init}=40\,M_\odot$ reaches the mass of Var\,2 ($M\approx23\,M_\odot$) and surface hydrogen abundance ${\rm X}_{\rm H}\approx40\%$ at age $t_{\rm age}\approx 4.8\,$Myr, its luminosity $L\approx4.8\times10^5\,L_\odot$ and effective temperature $T_{\rm eff}\approx15\,$kK are notably lower than those of Var\,2, while with $M_{\rm init}=60\,M_\odot$, star attains the mass of Var\,2 only when it has already evolved into a carbon-sequence Wolf-Rayet (WC) star with $L\approx8.0\times10^5\,L_\odot$ and $T_{\rm eff}\approx80\,$kK at age $t_{\rm age}\approx 4.0\,$Myr. Evolutionary tracks with moderate rotation rate $v_{\rm rot}/v_{\rm crit}=0.4$ (Figure~\ref{fig:evol_tracks}, right panel) show that star with $M_{\rm init}=40\,M_\odot$ will have luminosity and temperature identical to those derived for Var\,2 only directly after the main sequence stage, and it will not evolve into the WR region after the red supergiant stage, which restrict this evolutionary path for the Var\,2, located to the left side of the LBV minimum instability strip \citep{Groh2009_hrcar}. Thus, we conclude that Var\,2 most likely occupies an intermediate position between evolutionary pathways of slowly rotating stars with initial stellar masses of 40\,$M_\odot$ and 60\,$M_\odot$, matching a star with $M_{\rm init}\approx50\,M_\odot$ and $t_{\rm age}\approx 4.4\,$Myr that is evolving from lower temperatures towards the WR stars.

\subsection{Wind properties}

The optical depth of the wind $\tau_{\rm W}=\int_{r_{\rm c}}^\infty \kappa_{\rm F} \rho \, {\rm d}r$ in the self-consistent model of the extended atmosphere of Var\,2 is at the level of $\tau_{\rm W}\approx 2$, which indicates that the wind has already passed the transition phase and in the present state is closer to the winds of late-WN stars than O-stars \citep{Lamers2002, Vink2011, Bestenlehner2014}. As shown by \cite{Vink2012}, the mass loss rate $\dot{M}_{\rm trans}$ corresponding to the transit from O- to WR-stars can be determined from the relationship between the wind efficiency parameter $\eta$ and the optical depth of the wind $\tau_{\rm W}$:

\begin{equation}
\eta \equiv \frac{\dot{M} v_\infty}{L/c}=f_{\rm corr} \tau_{\rm W},
\end{equation}
where $f_{\rm corr}\simeq(\Gamma-1)/\Gamma$ -- correction multiplier expressed through the Eddington factor $\Gamma=g_{\rm rad}/g_{\rm grav}$. Thus, adopting optical depth of the wind in the resulting model $\tau_{\rm W}\approx2$ and the transitional optical depth $\tau_{\rm W}=1$, the transition mass-loss rate $\dot{M}_{\rm trans}\approx 10^{-5}\,M_{\odot}\,\text{yr}^{-1}$ for Var\,2 is at a level approximately twice lower than the observed one. The correction factor $f_{\rm corr}\approx0.2$ obtained for the wind of Var\,2 is well below the value $f_{\rm corr}=0.6$ derived by \cite{Vink2012} for the most massive stars in the Arches cluster.

As shown in Section~\ref{hydro_section}, the baseline Var\,2 wind model with $\beta=1.0$ underestimates the wind acceleration of the star. The calculation of a self-consistent velocity structure with higher wind acceleration beyond the sonic point increases the contribution of optically thick lines to the total radiation pressure, following the dependence $g^{\rm thick}_{\rm rad}\propto \frac{{\rm d}v}{{\rm d}r}/{\rho r^2}$. Considering that the radiative acceleration in both the continuum ($g^{\rm cont}_{\rm rad}\propto 1/r^2$) and optically thin lines ($g^{\rm thin}_{\rm rad}\propto 1/r^2$) is independent of the velocity gradient, one can estimate the effective ratio of the acceleration in optically thick lines to the total radiative acceleration $\alpha=g^{\rm thick}_{\rm rad}/g^{\rm tot}_{\rm rad}$. This ratio is used to parameterize the line pressure in the classical theory of stellar winds, where $\alpha$ is known as one of the line-force multiplier parameters \citep{Castor1975, Abbott1982, Gayley1995, Puls2000}. In the transonic wind region, the derived values of $\alpha$ fall within the range $\approx0.06-0.25$, indicating a relatively moderate dependence of radiative acceleration on the velocity gradient.

We can conclude that the Var\,2 wind, in terms of the ratio $g^{\rm thick}_{\rm rad}/g^{\rm tot}_{\rm rad}$, occupies an intermediate position between the optical thin winds of O-stars with $\alpha\approx0.5-0.7$ \citep{GormazMatamala2019, GormazMatamala2022, Poniatowski2022} and the densest winds of WR-stars, for which $\alpha \rightarrow 0$ \citep{GrafenerHamann2005}. A similar result was obtained by \cite{Grafener2008} for late-WN stars, which share many observational characteristics with certain LBVs/cLBVs \citep{Humphreys2014, Humphreys2017}. As was noted by \cite{GrafenerHamann2005}, such values of $\alpha$ presumably not reflect the original distribution of lines in the wind, but rather indicates the strong influence of numerous radiative effects (e.g. emission cascades due to recombination or extreme line overlaps) on the value of $\alpha$.

As seen in Figure~\ref{fig:hydro_vel} (blue line), the wind acceleration zone of the obtained self-consistent model can be roughly divided into two parts. These regions correspond to different ionization states of the iron group elements -- the main contribution to the wind acceleration in the inner and outer parts of the extended atmosphere is made by Fe\,IV and Fe\,III ions, respectively. Because the gas in the extended atmosphere of the star is in such a transition state, even minor variations in the model parameters can strongly alter the ionization balance of the wind. Note that Fe\,IV remains the dominant iron ion in the inner wind regions for all models presented above. In contrast, the stars crossing the bi-stable limit exhibit different behavior: recombination of Fe\,IV ions into Fe\,III occurs near the critical point \citep{Vink1999, Groh2011}, significantly affecting the mass-loss rate \citep{Pauldrach1990, Petrov2016} and, thus, playing a key role in the evolution of massive stars \citep{Sabhahit2022, Vink2023}.

The two-component velocity profile of the self-consistent wind model cannot be fully reproduced by the standard power-law distribution. In order to accurately describe the behavior of the Var\,2 wind velocity, it is necessary to use a more complex form of the $\beta$-law with two acceleration zones \citep{Hillier1999, GrafenerHamann2005}. A rough approximation of the wind velocity in the hydrodynamic model by a simple analytical law with different $\beta$ values (expression \ref{betalaw}) showed that the wind acceleration is best reproduced by $\beta\approx0.8$. The obtained velocity distribution is closer to the one of O- and early B-supergiants $\beta \approx0.5-1.5$ \citep{Vink2018, Sundqvist2019} than to the velocity profiles of WNL stars $\beta\approx1.4-1.7$ \citep{Grafener2008, Lefever2025}, while the ratio of the wind momentum $\dot{M} v_\infty$ to the luminosity of Var\,2 is in good agreement with the winds of WNL stars \citep{Crowther1995, Crowther1997, Sander2014}. At the same time, Var\,2 has a lower terminal wind velocity $v_\infty\approx230\,$km\,s$^{-1}$ compared to typical early-type supergiants or WNL stars (see references above), which is probably due to the proximity of LBVs to the Eddington limit \citep{Humphreys1994, Humphreys2014}.

There are no spectra of Var\,2 available outside the optical range, which restricted the determination of clumps stratification, a property tightly coupled to the velocity structure of the wind. Therefore, in our models we adopted a clumping parametrization derived from multi-wavelength studies of other massive stars, applying only minor corrections during hydrodynamic calculations to match the observed wind velocity. Future validation of such self-consistent non-LTE models for the slow and dense winds of LBVs, in terms of both velocity and inhomogeneities distributions in the wind, could be performed using spectral data covering a wider wavelength range.

\subsection{Stellar environment}\label{stellar_env_section}

\begin{figure*}[t]
     \centering
         \includegraphics[width=0.5\linewidth]{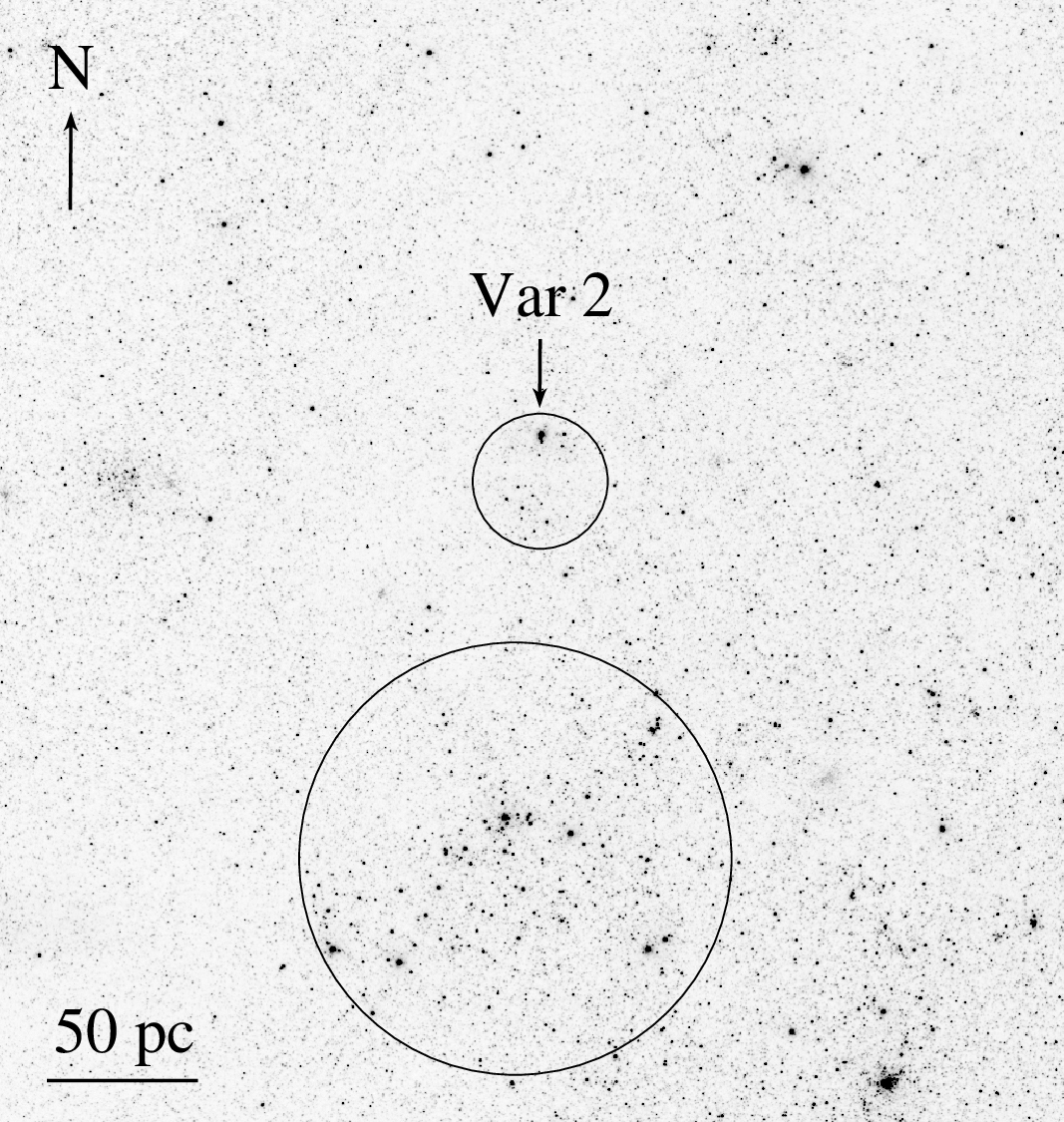}
        \caption{$F814W$-band image of the galaxy M33 field near Var\,2, obtained with the ACS/WFC camera on the HST. The circles mark stellar associations with ages estimated from color-magnitude diagrams.
        }
        \label{fig:var2_env}
\end{figure*}

\begin{figure*}[t]
     \centering
         \includegraphics[width=0.60\linewidth]{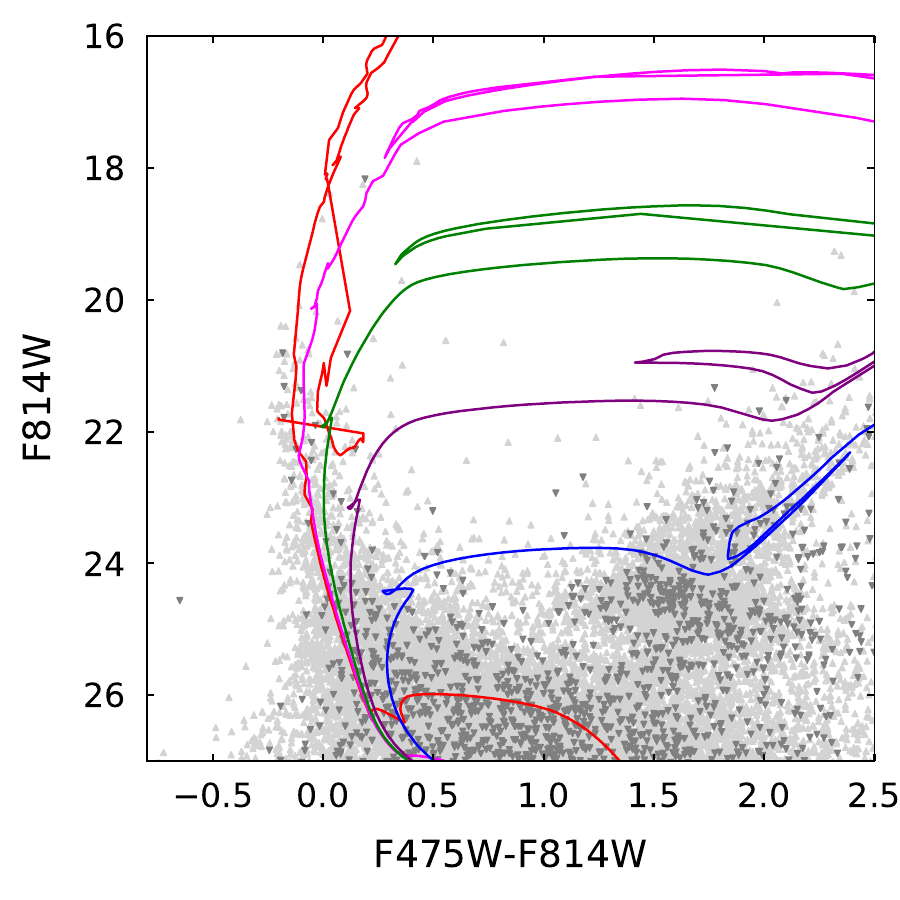}
        \caption{Color-magnitude diagram for stars in the regions marked in Figure~\ref{fig:var2_env}. Light- and dark-gray triangles denote stars from the larger and smaller regions. Theoretical isochrones from the {\tt PARSEC} evolutionary code for ages of 5, 10, 30, 100, and 400 Myr are shown as red, pink, green, purple, and blue curves, respectively.}
        \label{fig:cmd}
\end{figure*}

\cite{Kostenkov2025} had shown that the spatial distributions for various samples of stars do correlate with the masses of the objects under consideration: more massive stars, regardless of their evolutionary status, are, on average, located significantly closer to compact groups of bright blue stars. This result was obtained by studying the projected distances of stars  identified by independent methods to the nearest centers of OB associations in the galaxy M33. According to \cite{Kostenkov2025}, the spatial distribution of the sample of LBVs and LBV candidates (cLBVs) is close to that shown by massive stars with $M_{\rm init}\gtrsim20\,M_\odot$ and WR stars. However, some cLBVs, which were classified as stars with Fe\,II emission lines according to their spectra \citep{Humphreys2017}, turned out to be significantly isolated relative to OB associations. Likewise, studies of LBVs/cLBVs with predominantly low temperatures $T_{\rm eff}\approx10\,$kK found in galaxies of the Local Volume showed a discrepancy between the ages of the LBVs/cLBVs and the stars surrounding them \citep{Solovyeva2020, Solovyeva2021, Solovyeva2023}. The results of the analysis of the kinematics of LBVs/cLBVs in the Local Group \citep{Chentsov2021, Aghakhanloo2022, Deman2024} also notably differ, suggesting that the kinematic properties may depend on the specific subtypes of the stars. All this points to the importance of simultaneously analyzing both the spectral features and the nearby stellar population of the LBVs. Below, we will present a detailed analysis of the stellar environment of Var\,2.

As shown in Figure~\ref{fig:var2_env}, Var\,2 is located at a projected distance of about 100 pc from a large star-forming region. There are also several fainter blue stars in the local neighborhood of this LBV. To study the stellar populations in the two regions marked in Figure~\ref{fig:var2_env}, we used the photometric catalog of the galaxy M33 based on the multi-band PHATTER survey \citep{Williams2021} and constructed color-magnitude diagrams (Figure~\ref{fig:cmd}). We determined the ages of the selected stellar groups using isochrones computed with the {\tt PARSEC} evolutionary code\footnote{https://stev.oapd.inaf.it/cgi-bin/cmd} \citep{Bressan2012}. The isochrone calculations adopted a reddening value $A_V=0.54$ ($R_V=3.39$; \citealt{Wang2022}) as derived in Section~\ref{modeling_section}.

The position of the youngest stellar population is best fitted by the isochrone with age of $t_{\rm age}\approx5\,$Myr, consistent with age estimate of Var\,2 $t_{\rm age}\approx4.4\,$Myr, implicitly confirming single massive star evolution scenario. If Var\,2 originally formed in a distant large star-forming region, its minimum projected velocity relative to the nearest blue star group would be $v_{\rm proj}\approx20\,$km\,s$^{-1}$. This estimate agrees well with both typical observed field OB-star velocities ($v_{\rm proj}\approx10\,$km\,s$^{-1}$; \citealt{Gies1987,Guo2024}) and cluster ejection velocities of massive O-stars ($v_{\rm ej}\approx10-20\,$km\,s$^{-1}$) from N-body simulations \citep{Oh2015,Oh2016}.

Consequently, Var\,2 could potentially have been formed in either of the two selected stellar groups. However, as shown in Section~\ref{obs_section}, there are two relatively bright blue stars in the immediate vicinity of Var\,2. This spatial configuration suggests Var\,2 likely belongs to this sparse group, as the probability of finding three massive stars randomly projected within the circle of radius $\approx0.6\arcsec{}$ area is low. To test this hypothesis, we calculated the probability of this spatial distribution under the assumption of Poisson statistics, where the probability of finding $n$ stars within radius $r$ is given by: 

\begin{equation}
P_n(r) = \frac{u^n}{n!} \exp\left({-u}\right),
\end{equation}
where $u=\pi r^2 \delta$ is the mean number of stars with a projected surface density $\delta$ within a circle of radius $r$. The density $\delta$ can be expressed in terms of the mean separation $\langle d \rangle$ between nearest-neighbor stars in the sample \citep{Ivanov1996}:

\begin{equation}
\delta\approx\left(\frac{3}{2  \langle d \rangle}\right)^2.
\end{equation}

The statistical probability of the observed spatial distribution was calculated using a sample of stars hotter than spectral type B0 with $M_{\rm bol}\lesssim-7.0,$mag (bolometric corrections for these spectral classes are ${\rm BC}\lesssim-3.0\,$mag; \citealt{Bedin2005}), which roughly matches the luminosity of stars surrounding Var\,2. Using the photometric catalog from \cite{Williams2021}, we selected such stars using the criteria $m_{\rm F475W}<21\,$mag and $(m_{\rm F336W}-m_{\rm F475W})<-1.0\,$mag. We calculated the projected density of selected OB stars within a $100\arcsec{}$ radius around Var\,2. The probability of finding three blue stars with $M_{\rm bol}\lesssim-7.0\,$mag within a $r=0.6\arcsec{}$ circle is $P_3(0.6\arcsec{})\approx10^{-4}$. 

The low probability leads us to the intriguing idea that Var\,2 likely formed in a sparse stellar group near its current location, rather than being ejected from some distant cluster associated with a large star-forming region. To date, a fairly large number of examples of massive stars located far from clusters or large star-forming regions are known (e.g. \cite{ZarricuetaPlaza2023, Kalari2019, Selier2011, Renzo2019}), however, the question of whether they are all "runaway"{} stars ejected from clusters, or some of them were formed "in situ"{}, is open (\cite{VargasSalazar2023}; see discussion and references in \cite{Gvaramadze2012}). Current theories suggest the possibility of the formation of massive stars both in the centers of clusters \citep{Bonnell2002, Bonnell2006, Peters2010} and in isolated massive clumps of gas, similar to low-mass stars, but with a significantly higher accretion rate \citep{Yorke2002, McKee2003, Krumholz2008}. Further study of high-luminosity stars outside clusters may improve our understanding of possible mechanisms for the formation of such objects.

\section{Conclusions}

The current research presents the first self-consistent non-LTE spectral modeling of a confirmed LBV, using Var\,2 as a case study. This LBV remains in a "dormant"{} state, showing no notable photometric variability for the past several decades following a gradual decline in brightness after its peak in the 1920s. Given the limited sample of LBVs with reliably determined parameters, this study represents a significant advancement in the understanding nature of LBVs and similar evolved massive stars.

The values of the current mass $M_* \approx 23\,M_\odot$, luminosity $L=6.5\times10^5\,L_\odot$ and the hydrogen abundance on the surface ${\rm X}_{\rm H}\approx43\%$, obtained for Var\,2 during modeling, match a star with the initial mass $M_{\rm init}\approx50\,M_\odot$ and age $t_{\rm age}\approx4.4\,$Myr. This point on the evolutionary trajectory corresponds to the transition of evolved massive star to the Wolf-Rayet phase, which agrees with the classical Conti scenario. Furthermore, the wind properties of Var\,2 also confirm its intermediate evolutionary state. The age of Var\,2 is consistent with ejection from a cluster $\sim$100 pc away, but the statistics of its stellar environment suggest it instead formed in a local low-populated group.

\section*{Acknowledgments}
We thank referee for valuable comments and John Hillier for providing the CMFGEN code. The work was performed as part of the SAO RAS government contract approved by the Ministry of Science and Higher Education of the Russian Federation. Observations with the SAO RAS telescopes are supported by the Ministry of Science and Higher Education of the Russian Federation. The renovation of telescope equipment is currently provided within the national project "Science and Universities". Based on data retrieved from the SOPHIE archive at Observatoire de Haute-Provence (OHP), available at atlas.obs-hp.fr/sophie.

\bibliographystyle{raa}
\bibliography{ms2025-0404}

\end{document}